\documentclass[aps,pra,twocolumn,amsmath,superscriptaddress]{revtex4-1}

\newcommand{\bea}{\begin{eqnarray}}
\newcommand{\eea}{\end{eqnarray}}
\newcommand{\beq}{\begin{equation}}
\newcommand{\eeq}{\end{equation}}

\newcommand{\ep}{\epsilon}

\newcommand{\si}{\sigma}
\newcommand{\Si}{\Sigma}

\usepackage[urlcolor=blue,colorlinks=true,citecolor=blue,linkcolor=blue,pdfstartview={FitH},bookmarks=false]{hyperref}
\usepackage{graphicx}
\usepackage{longtable}
\usepackage{epsfig}
\usepackage{dcolumn}
\usepackage{bm}
\usepackage{amssymb}
\usepackage{multirow}
\usepackage{times,color}
\usepackage{hyperref}

\begin{document}

\title{Quantum phase transitions in the spin-1 Kitaev-Heisenberg chain}

\author{Wen-Long You}
\affiliation{College of Science, Nanjing University of Aeronautics and Astronautics, Nanjing, 211106, China}
\affiliation{School of Physical Science and Technology, Soochow University, Suzhou, Jiangsu 215006, China}

\author{Gaoyong Sun}
\affiliation{College of Science, Nanjing University of Aeronautics and Astronautics, Nanjing, 211106, China}

\author{Jie Ren}
\affiliation{Department of Physics, Changshu Institute of Technology, Changshu 215500, China}

\author{Wing Chi Yu}
\affiliation{Department of Physics, City University of Hong Kong, Kowloon, Hong Kong }

\author{Andrzej M. Ole\'s$\,$}
\affiliation{\mbox{Institute of Theoretical Physics, Jagiellonian University,
             Profesora Stanis\l{}awa \L{}ojasiewicza 11, PL-30348 Krak\'ow, Poland}}
\affiliation{\mbox{Max Planck Institute for Solid State Research,
             Heisenbergstrasse 1, D-70569 Stuttgart, Germany} }

\begin{abstract}
Recently, it has been proposed that higher-spin analogues of the Kitaev
interactions $K>0$ may also occur in a number of materials with strong
Hund's and spin-orbit coupling.  
In this work, we use Lanczos diagonalization and density matrix
renormalization group methods to investigate numerically the $S=1$
Kitaev-Heisenberg model.
The ground-state phase diagram and quantum phase transitions
are investigated by employing local and nonlocal spin correlations.
We identified two ordered phases at negative Heisenberg coupling $J<0$:
a~ferromagnetic phase with $\langle S_i^zS_{i+1}^z\rangle>0$ and
an intermediate left-left-right-right phase with
$\langle S_i^xS_{i+1}^x\rangle\neq 0$. A~quantum spin liquid is stable
near the Kitaev limit, while a topological Haldane phase is found
for $J>0$.
\end{abstract}

\date{20 August, 2020}

\maketitle

Kitaev-Heisenberg (KH) models were fostered by an endeavor of achieving
the Kitaev physics in transition metal oxides~\cite{Nus15}.
A continuing interest of bond-directional interactions is motivated by
topological quantum computing \cite{Aas20}, especially after Kitaev 
proposed an exactly solvable model of frustrated quantum spins $S=1/2$ 
on a two-dimensional (2D) honeycomb lattice with bond-directional 
interactions \cite{Kitaev}. The Kitaev model was initially treated as a 
mathematical model describing a topological quantum spin liquid (QSL) 
ground state (GS) and Majorana excitations, until Jackeli and Khaliullin
\cite{Jackeli09} demonstrated that the bond-directional interactions
could be realized in Mott insulators with strong spin-orbit coupling.
This innovative concept initiated intense theoretical and experimental
search for the $S=1/2$ Kitaev QSLs in solid state materials
\cite{Dag17}. It has been found that other interactions, such as the
isotropic Heisenberg and/or off-diagonal exchange terms contribute
\cite{Rau14,Juraj,Wan20,Ian}, and real systems do not realize the QSL.

The 2D model appears difficult to analyze and its
phase diagram has a QSL in the Kitaev limit \cite{Cha13,Got17}, but even
the one-dimensional (1D) version of it has several interesting quantum
phase transitions (QPTs) \cite{Agr18}. A spin-1/2 1D variant of KH model 
was defined on a chain, in which two types of nearest-neighbor Kitaev 
interactions sequentially switch between $S_i^xS_{i+1}^x$ on odd and 
$S_i^yS_{i+1}^y$ on even bonds next to uniform Heisenberg
interactions. The GS phase diagram of spin-$1/2$ KH model was depicted
using the density matrix renormalization group (DMRG) and exact
diagonalization (ED) methods \cite{Agr18}. Much attention 
has been paid to the Kitaev limit
\cite{Brzezicki07,Brzezicki09,You08,You10,You11,Eriksson,Subrahmanyam,Liu}.
The two-spin correlation functions are found to be extremely short-ranged
\cite{Baskaran07,You12}, indicating a QSL state.

Recently it was realized that $S\!=\!1$ KH model could be designed by
considering strong Hund's coupling among two electrons in $e_g$ orbitals
and strong spin-orbit coupling (SOC) at anion sites~\cite{Stavropoulos19}.
However, relatively little is known about the magnetic properties and
particularly the elementary excitation spectrum for higher $S$ the
theoretical investigation on the effect of Heisenberg exchange in the
Kitaev chain. It has been realized long after Haldane's pioneering
work~\cite{Haldane1,Haldane2}  that spin models with integer or half-odd
integer $S$ are qualitatively different. The N\'{e}el state is favored by
Heisenberg antiferromagnetic (AFM) term for half integer spin $S$,
while cannot play a similar role when $S$ is an integer.

It is recognized that the GS of the $S=1$ Heisenberg
antiferromagnet belongs to the Haldane phase, which is separated from
all excited states by a finite spin gap~\cite{White93}, and thus
two-spin correlation is quenched. The underlying physics of Haldane
chains is fairly well understood both in theory and experiments. The
Haldane phase of spin-1 XXZ AFM chains was proposed in trapped ions
systems~\cite{Cohen14}. For instance, a hidden
$\mathbb{Z}_2\times\mathbb{Z}_2$ symmetry breaking takes place
\cite{Kennedy,Takada} and hence the string order parameters are nonzero
in both $x$- and $z$-directions \cite{Nijs,Tasaki}. When the Kitaev
interaction is taken into account, the spin chains only have a
$\mathbb{Z}_2$ parity symmetry corresponding to the rotation of $\pi$
around a given axis~\cite{Berg}. If the $\mathbb{Z}_2$ symmetry in the
GS is broken, whether the string order parameter along
the given axis becomes nonzero is unclear~\cite{Hatsugai}. Therefore,
it is also an interesting issue to explore the existence of the string
correlators in spin-1 chains with lower symmetries than Heisenberg chain.
The phase diagram of spin-1 generalized Kitaev chain (also dubbed as
compass model in the literature) were also investigated \cite{Liu15}.
The GS properties and the low-energy excitations of spin-1 KH models
are elusive and deserve a careful investigation.

\begin{figure}[b!]
\includegraphics[width=\columnwidth]{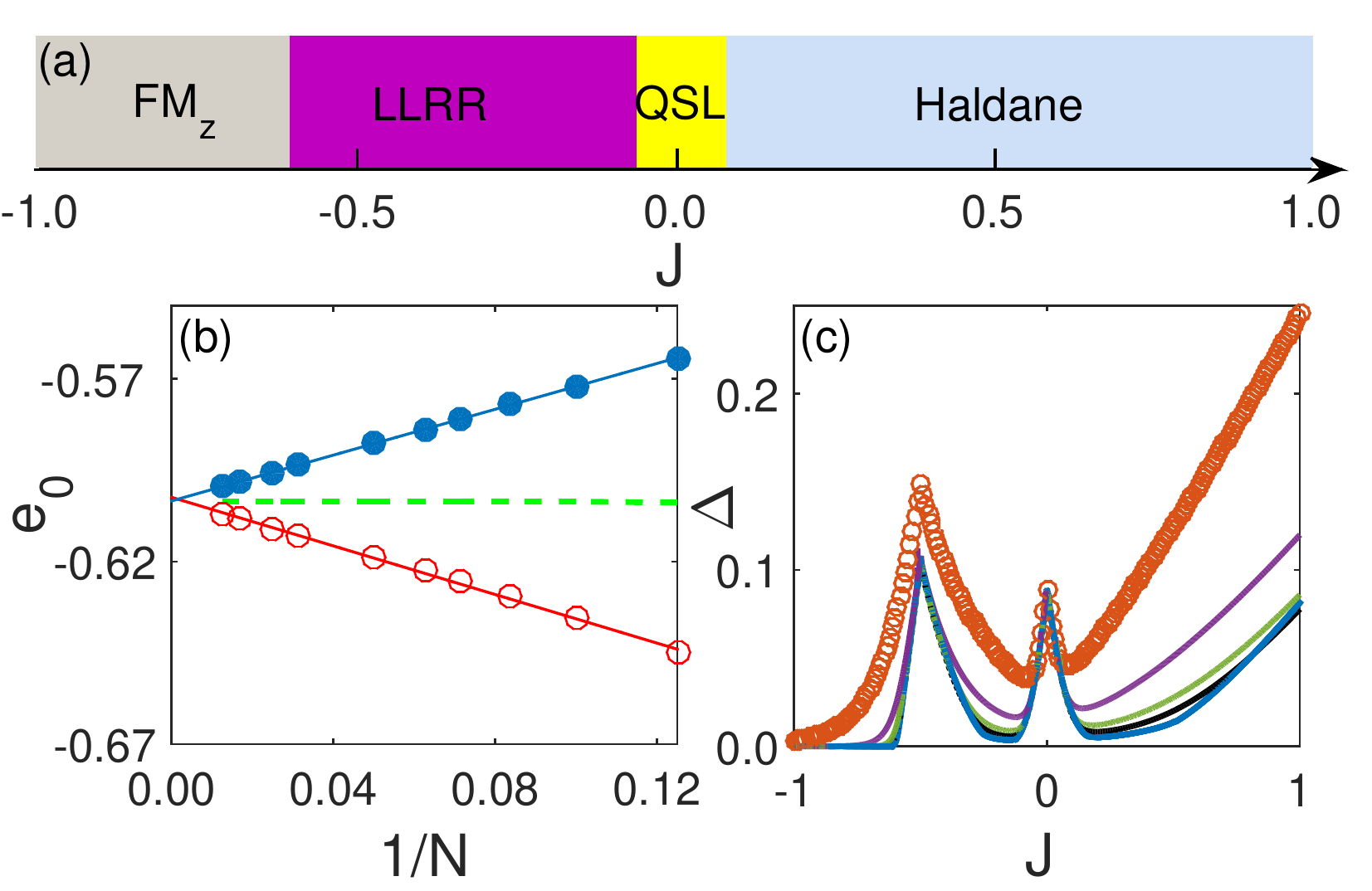}
\caption{(a) Schematic phase diagram of the KH model with $K=1$;
(b) finite size scaling of the energy density for OBC and $J=0$;
(c) the energy gap for system sizes ranging from $N$=12 to $N$=100.
The blue and red symbols in (b) represent the GS energy of $\hat{H}_K$
per site and per bond, respectively. The linear fits correspond to
\mbox{$e_0(N)\!=-0.3326/N-0.6024$} and
\mbox{$e_0(N)\!=-0.3136/N-0.6034$.}
The green dashed line is $e_0(\infty)$.
}
\label{fig1}
\end{figure}

The purpose of this paper is twofold. First, we would
like to obtain the GS phase diagram and discuss the QPTs in the 1D
spin-1 KH model. Second, while some differences in the structure of the
invariants between the models with half-odd integer and integer spins
have been pointed out~\cite{Bas08}, the issue of whether there are
systematic differences in the nature of the low-energy spectrum is open
\cite{Sen10}. The main result of our study is that the GS of the spin-1
KH chain with periodic boundary conditions (PBCs) changes from the QSL
to the left-left-right-right (LLRR) (Haldane) phase for $J\!<\!0$
($J\!>\!0$), see Fig. \ref{fig1}(a). Both phases are unique for the
$S=1$ 1D KH model and we employ the ED and the DMRG. In the DMRG
simulations, we keep up to $m=500$ eigenstates during the basis
truncation and the number of sweeps is $n=30$. These conditions
guarantee that the simulation is converged and the truncation error is
smaller than $10^{-7}$.

In the present paper we deal with a spin-1 KH chain,
\begin{eqnarray}
\label{Hamiltonian}
\hat{H} &=& \hat{H}_{K} + \hat{H}_{J},\\
\hat{H}_{K}&=&K\sum^{N/2}_{j=1}\left(S^x_{2j-1}S^x_{2j}+S^y_{2j}S^y_{2j+1}\right)
\label{Ham1},\\
\hat{H}_{J}&=&J \sum^{N}_{j=1} {\bf S}_j \cdot{\bf S}_{j+1}.
\label{Ham2}
\end{eqnarray}
Here ${\bf S}_j=\{S_{j}^x,S_{j}^y,S_{j}^z\}$ are the spin-1 operators
at site $j$, and $N$ is the total number of sites. The parameters
$\{K,J\}$ stand for the Kitaev and Heisenberg exchange coupling.
Hereafter, we set \mbox{$K\equiv 1$.} We deal with spin-1 operators in
a special representation, $S^\alpha_{bc}=i\ep_{abc}$, i.e.,
$\{S^x,S^y,S^z\}$ are given by:
\begin{eqnarray}
\left(
\begin{array}{ccc}
0 & 0 & 0 \\
0 & 0 & -i \\
0 & i & 0 \\
\end{array}
\right),\quad
\left(
\begin{array}{ccc}
0 & 0 & i \\
0 & 0 & 0 \\
-i & 0 & 0 \\
\end{array}
\right),\quad
\left(
\begin{array}{ccc}
0 & -i & 0 \\
i & 0 & 0 \\
0 & 0 & 0 \\
\end{array}
\right).
\label{s1adef}
\end{eqnarray}
Spin operators $\{S_j^\alpha\}$ at site $j$ in Eq.
(\ref{Hamiltonian}) obey the SU(2) algebra,
$[S_i^\alpha,S_j^\beta]=i\delta_{ij}\epsilon_{\alpha\beta\gamma}S_j^c$,
with the
totally antisymmetric tensor $\epsilon$ and $({\bf S}_j)^2=S(S+1)=2$.

%
First we consider the Kitaev limit in Eq. (\ref{Hamiltonian}), i.e.,
$J=0$. Then the global spin rotation SU(2) symmetry is not conserved.
We can write the spin operators in $\hat{H}_{K}$ in terms of the ladder
operators $S_j^\pm\equiv S_j^x\pm i S_j^y$, and one finds that
$[S_j^z,S_j^\pm]=\pm S_j^\pm$, i.e., the Ising terms in Eq. (\ref{Ham1})
change the total pseudospin $z$th component at both odd $x$-link and
even $y$-link by either 0 or $\pm 2$. A site parity operator is
$\Si^\alpha_j=e^{i\pi S_j^\alpha}$, i.e., $\{\Si^x,\Si^y,\Si^z\}$ are
given by the diagonal matrices that satisfy
$\Si^\alpha_j\!=\!1\!-\!2(S_j^\alpha)^2$ and $\Si_j^x\Si_j^y\Si_j^z=I$,
where $I$ is an identity matrix. The Hamiltonian in Eq. (\ref{Ham1})
has a global discrete symmetry with respect to rotation by an angle
$\pi$ about the $x$, $y$, $z$ axes, i.e., $\prod_j \Si_j^\alpha$,
present in the dihedral group $D_2$. The time reversal symmetry, i.e.,
$S_j^{x,y,z}\to -S_j^{x,y,z}$, and the spatial inversion symmetry,
i.e., $S_j^{x,y,z}\to S_{N+1-j}^{x,y,z}$, are also respected.

Furthermore, one finds all $\{\Si^\alpha_j\}$ matrices commute with
each other. In addition, $\Si^\alpha_j$ commutes with $S_j^\alpha$ but
anticommutes with $S_j^\beta$ ($\alpha\!\neq\!\beta$), i.e.,
$\{\Si^\alpha_j,S_j^\beta\}\!=\!\{\exp(i\pi S_j^\alpha),S_j^\beta\}\!=
\!0$. In this regard, the bond parity operators on odd/even bonds,
\begin{eqnarray}
W_{2j-1}=\Si^y_{2j-1} \Si^y_{2j}, \quad\quad
W_{2j}=\Si^x_{2j} \Si^x_{2j+1},
\end{eqnarray}
define the invariants of the Hamiltonian in Eq. (\ref{Ham1})
and eigenvalues of $W_j$ are $\pm 1$. It can be verified that
 $[W_j, W_k]=0$,   
 $[W_j, \hat{H}_{\rm K} ]=0$.
The GS of $\hat{H}_{\rm K}$ lies in the sector with all $W_j=1$ which
can be proved by applying the reflection positivity technique in the
spin-1/2 counterpart~\cite{You08}. In the GS sector, the system can be
mapped to a single qubit-flip model with nearest neighbor exclusion
represented by the effective Hamiltonian \cite{Sen10}:
\beq
\tilde{H}_{\rm K,GS}=\frac14\sum_j
\left(1-\si^z_{j-1}\right)\,\si^x_j\,\left(1-\si^z_{j+1}\right).
\label{ham3}
\eeq
At \mbox{$J\!=\!0$} the spectrum is gapped ($\Delta>0$) and the first
excited state is $N$-fold degenerate, corresponding to one $W_j=-1$
defect in the $W_j=1$ sector. The energy gap is insensitive to the
system size, see Fig. \ref{fig1}(c), and remains finite in the
thermodynamic limit, $\Delta=0.1764$. Below we show that the Kitaev QSL
phase survives in a finite range of coupling $J$.

In the case of open boundary conditions (OBCs) the GS is fourfold
degenerate. We consider a finite system with the OBC in the ED method.
It is easy to show that for chains with OBCs the GS energy per site
\mbox{$e_0^l\equiv E_0(N)/N$} [per bond $e_0^u\equiv E_0(N)/(N-1)$]
gives a lower (upper) bound for the GS energy of an infinite
chain \cite{You07}. The $N$-dependence of the GS energy
$e_0(N)\!=\!E_0/N$ can be fitted to the form
$e_0(N)  = e({\infty}) + c/N$,
where $c$ is independent of $N$. Indeed, $\log|e_0(N)-e({\infty})|$
gives $e({\infty})\!\approx\! -0.6029$, see Fig. \ref{fig1}(b).

\begin{figure}[t!]
 \includegraphics[width=\columnwidth]{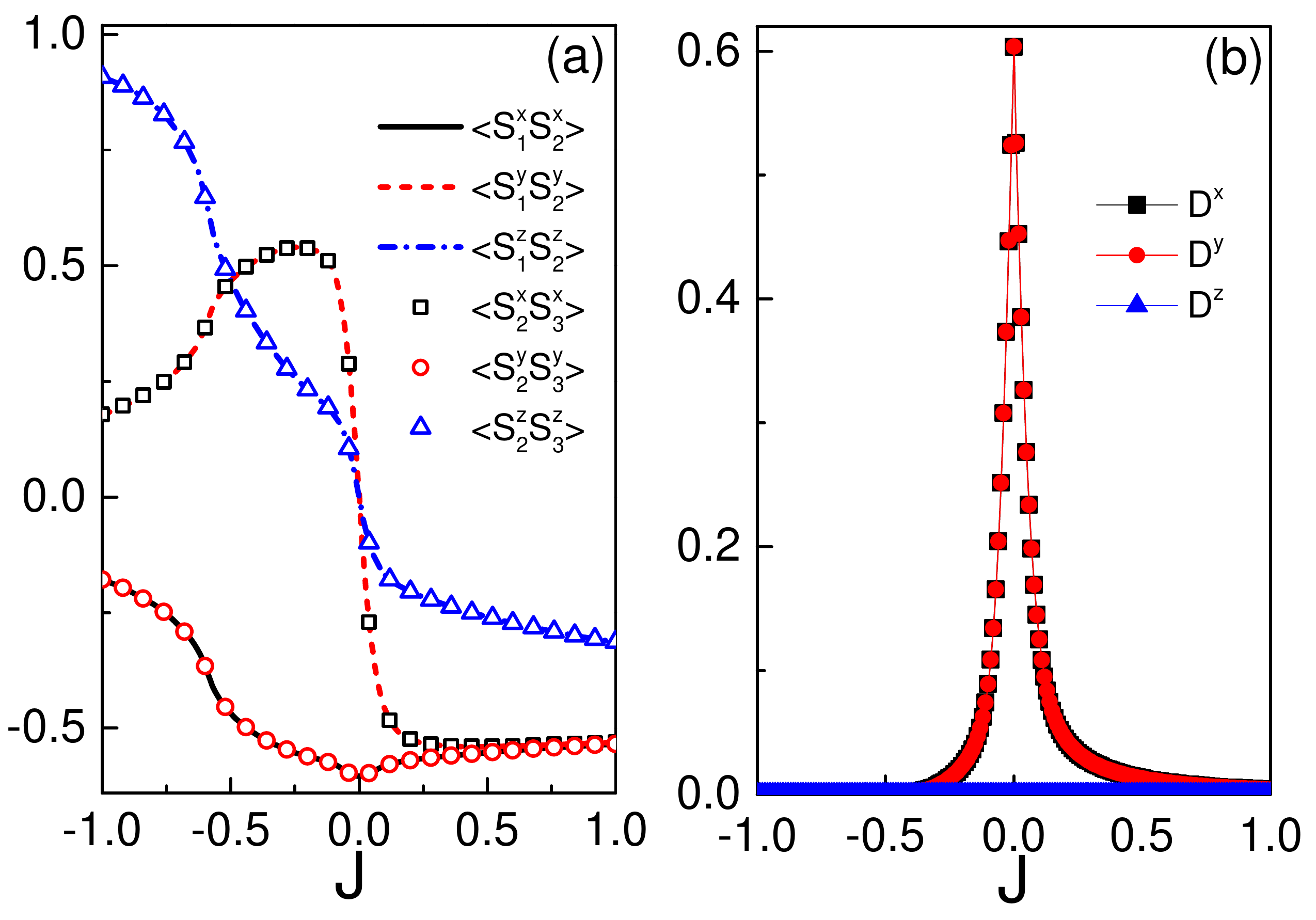}
\caption{(a) Two-point spin correlators $C^\alpha(i,j)$ (\ref{spin}) and 
(b) the dimer order parameter $D^\alpha$ (\ref{dimerOP}).
Parameters: increasing $J$ and $N=60$.}
\label{fig2}
\end{figure}

According to Ginzburg-Landau theory, a well defined order parameter is
a vital ingredient for characterizing the quantum phase. In order to
characterize the Kitaev phase and the QPTs at $J\neq 0$, we calculate
the two-point spin correlations,
\begin{eqnarray}
\label{two-point}
C^{\alpha}(i,j)=\langle S^{\alpha}_iS^{\alpha}_j\rangle,~\quad \alpha=x,y,z,
\label{spin}
\end{eqnarray}
where $i$ and $j$ specify the positions of the initial and final sites,
respectively. Accordingly, $r=|j-i|$ measures the separation between the
two sites. Similar to its spin-1/2 counterpart, the two-point
correlators precisely vanish beyond nearest neighbors due to the
$\mathbb{Z}_2$ symmetry, sharing many of the same properties and
phenomenology of the spin-1/2 Kitaev honeycomb model \cite{Dag17}.
The pure spin-1 Kitaev chain hosts only two nearest neighbor AFM orders
$C^x(2i-1,2i)$ on $x$-links and $C^y(2i,2i+1)$ on $y$-links, as is
shown in Fig. \ref{fig2}(a).

Next we turn on a nonzero perturbations $\propto J$, which could
undermine the conservation of local quantities characteristic of the
Kitaev model. We assume that such interactions are of Heisenberg type,
as suggested by possible solid state applications. We then make a
comprehensive study on the phase diagram of the KH Hamiltonian in its
full parameter space, using a combination of extensive analytical,
DMRG and ED calculations. It was reported that spin-1 Kitaev honeycomb
model in candidate materials, such as honeycomb Ni oxides with heavy
elements of Bi and Sb, is accompanied by a finite ferromagnetic (FM)
Heisenberg interaction. The Kitaev QSL is stable in a range of
$|J|/K>0.08$ \cite{Hickey20} and infinitesimal $J$ does not destabilize
it.

It is widely recognized that the GS has a qualitative difference between
an integer and half-odd integer spin-$S$ models. For $S=1/2$, the Kitaev
GS is $2^{N/2-1}$-fold degenerate and such macroscopic degeneracy makes
it fragile. Accordingly, an infinitesimal Heisenberg coupling is
sufficient to lift the GS degeneracy and to generate magnetic long-range
order in the compass-Heisenberg model, either FM or AFM one~\cite{Tro10}
when the Heisenberg interactions spoil the $\mathbb{Z}_2$ symmetry
associated with each bond. It is worth noting that the low-lying
excited-state energy level crossings at $J=0$ takes place, which plays
an analogous role in the $J_1$-$J_2$ Heisenberg chain~\cite{Chen07}.
Although the second-order derivative
of energy density and the normalized fidelity susceptibility exhibit a
local peak, the peak declines with increasing system size $N$.

The Hamiltonian is invariant under a rotation around the $z$-axis by an
angle $\pi/2$ (i.e., $S_j^x\to S_j^y$, $S_j^y\to -S_j^x$)
and a translation by one lattice site with $i\to i+1$. The combination
of rotation and translation symmetries imply that $C^x(1,2)=C^y(2,3)$,
$C^y(1,2)=C^x(2,3)$, and $C^z(1,2)=C^z(2,3)$, which are confirmed
in Fig. \ref{fig2}(a). A small Heisenberg coupling can induce other
correlations, especially such as
$\langle S^y_{2i-1}S^{y}_{2i}\rangle$ and
$\langle S^x_{2i}S^{x}_{2i+1}\rangle$. As shown in Fig. \ref{fig2}(b),
the bond-directional order in the Kitaev QSL phase can be captured by
the dimer order parameter,
\begin{equation}
\label{dimerOP}
D^\alpha=\left\vert\left\vert C^\alpha(2i-1,2i)\right\vert
                  -\left\vert C^\alpha(2i,2i+1)\right\vert\right\vert.
\end{equation}
We have verified that the finite-size effects are negligible.

\begin{figure}[t!]
\includegraphics[width=\columnwidth]{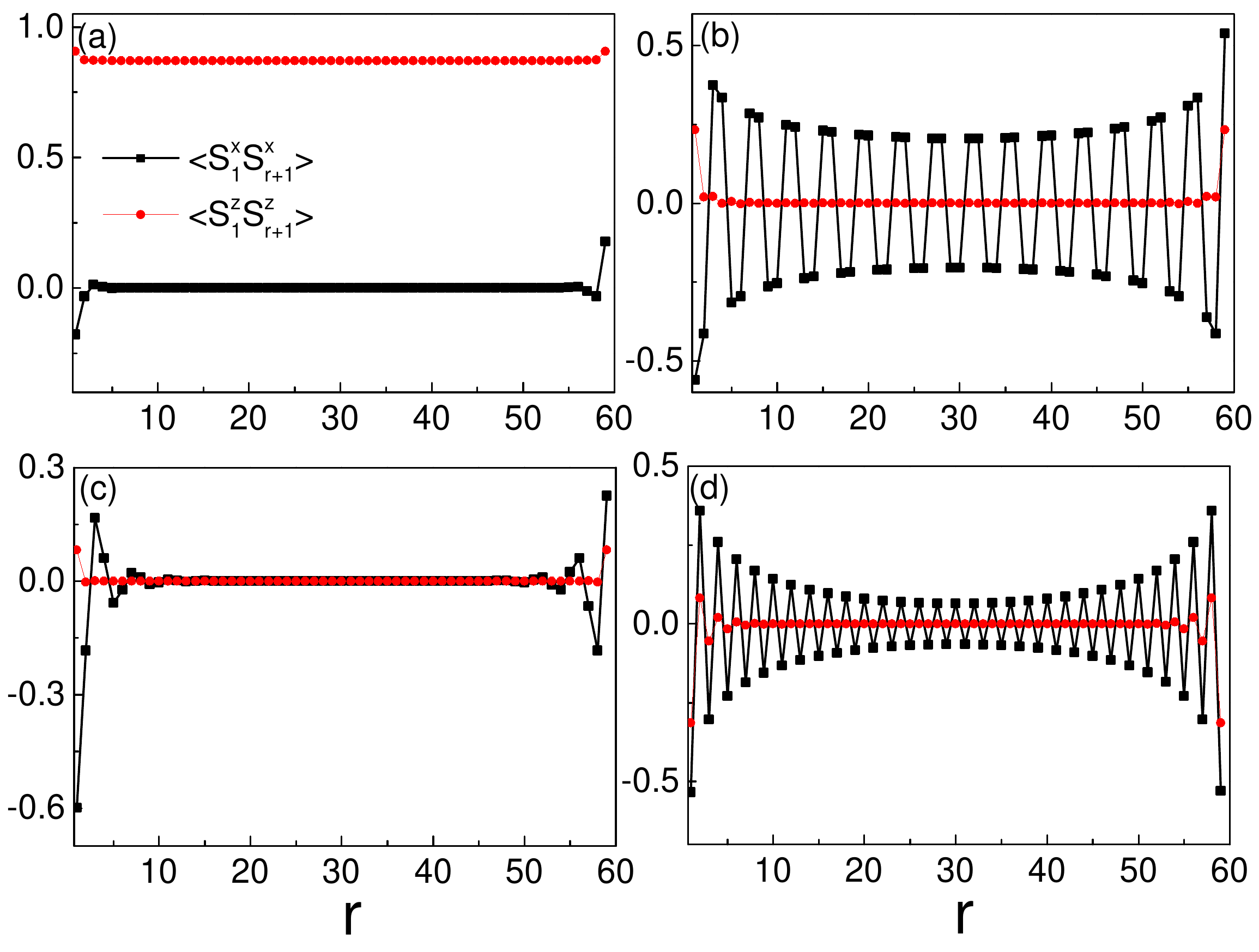}
\caption{The correlation between site 1 and site $i$ for increasing
distance $r\equiv \vert i-1 \vert $ for four representative points in
different phases (Fig. \ref{fig1}): (a) $J=-1$; (b) $J=-0.2$;
(c) $J=-0.03$, and (d) $J=1$. Here we use the PBCs for $N=60$.}
\label{fig3}
\end{figure}

When $J$ varies, the competing correlations will trigger miscellaneous
phase transitions. For FM Heisenberg exchange interaction $J<0$,
the dominating $x$-component correlations have a negative (positive)
sign on odd (even) bonds, evincing the system develops the LLRR spin
order, see Fig. \ref{fig2}(a). This is similar to the GS spin
configuration in the ANNNI model, in which the nearest neighbor
interactions favor the FM alignment of neighboring spins while
interactions between the next nearest neighbors foster
antiferromagnetism. For $J\simeq -1$, the $C^z(i,j)$ correlations
dominate and FM$_z$ GS is found, see Fig. \ref{fig3}(a). When $J$
increases from $J\!=\!-1$ to $J\!=\!0$, the chain undergoes two 
successive second-order QPTs at $J_{c1}\!\simeq\! -0.6$ and
$J_{c2}\!\simeq\! -0.08$ (Fig. \ref{fig1}). Figure \ref{fig3} confirms
that spin correlations are crucial and identify QPTs shown in Fig.
\ref{fig1}(a).

On the other hand, the GS of Eq. (\ref{Ham2}) with AFM couplings ($J>0$)
is a topological phase predicted by Haldane \cite{Yang20}, which has a
finite excitation gap $\Delta=0.41J$ and exponentially decaying spin
correlation functions. More precisely, since the edge states have a
finite length for an open chain, the splitting in the lowest energies is
exponentially small for longer chains, resulting in fourfold
quasidegenerate GSs below the Haldane gap~\cite{Kennedy90}. It is well
known that this phase cannot be characterized by any local
symmetry-breaking order parameter. In view of the analogy of GS
degeneracy of spin-1 Kitaev and Heisenberg models, a natural question is
whether the GS of the Kitaev chain can be adiabatically connected to
the Haldane phase without going through a phase transition. The
topological nature of the Haldane phase becomes especially clear after
Affleck, Kennedy, Lieb, and Tasaki (AKLT) proposed the exactly solvable
AKLT model~\cite{Affleck87}, whose GS exhibits intriguing properties,
such as a nonlocal string order and $2^2$ edge states composed of two
free $S=1/2$ spinons. Thus, we investigate the string order parameter
\cite{Nijs},
\begin{equation}
\label{Osigma}
O^\alpha(l,m)=\left\langle S^\alpha_l\exp\left(
i\pi\sum_{k=l+1}^{m-1}S_k^\alpha\right)S^\alpha_m\right\rangle,
\end{equation}
whose limiting value
$O^{\alpha}_s={\rm lim}_{|l-m|\rightarrow\infty}\{-O^{\alpha}(l,m)\}$,
reveals the hidden symmetry breaking, where the $\vert 1\rangle$
($\vert$-$1\rangle$) states alternate diluted by arbitrary strings of
$\vert 0\rangle$. Here $\vert m\rangle$ is an eigenstate of $S^\alpha$
with an eigenvalue $m=-1,0,1$.

\begin{figure}[t!]
\includegraphics[width=\columnwidth]{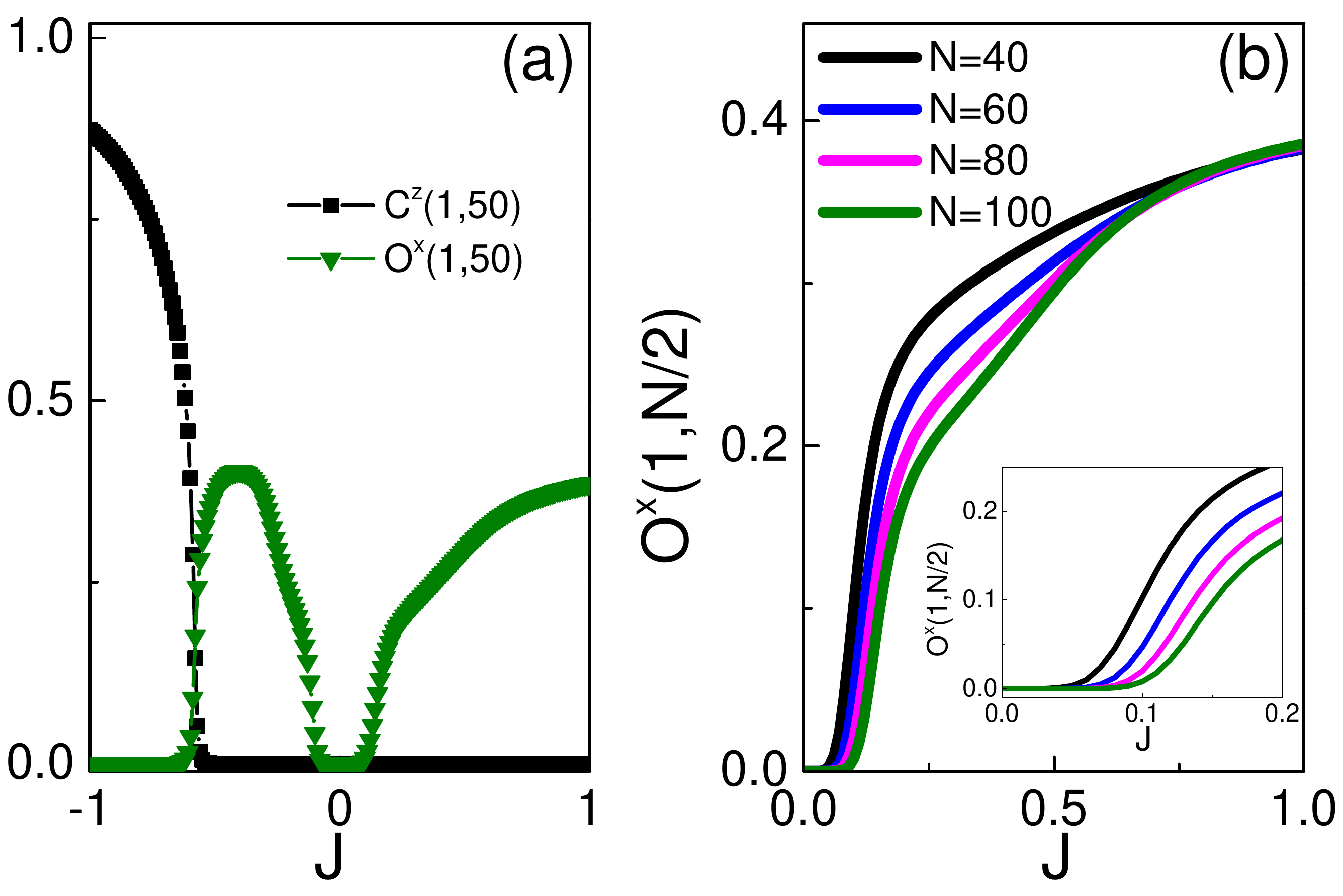}
\caption{(a) Two-point correlation $C^z$ (\ref{spin}) and the string
order parameter $O^x$ (\ref{Osigma}) between sites 1 and 50 for $N=100$
and $J\in(-1,1)$; 
(b) The string order parameter $O^x$ (\ref{Osigma}) between sites 1 and
$N/2$ in the Haldane phase ($J>0$). 
Inset amplifies the region close to $J=0$.}
\label{fig4}
\end{figure}

Applying the Kennedy-Tasaki (KT) transformation~\cite{Oshikawa92},
$U_{\mathrm{KT}}=\prod_{j<k}\exp{\left( i\pi S^z_j S^x_k \right)}$,
one transforms the diluted AFM phase into the phase containing only
$|0\rangle$ and $|1\rangle$ or only $|0\rangle$ and $|{\rm -}1\rangle$
states, and converts the nonlocal string order into the local FM order.
In this regard, Eq. (\ref{Ham2}) is transformed into a Hamiltonian with
short-range interactions,
\mbox{$\tilde{H}_{J}\!=-J\sum_i^N \left(S_i^xS^x_{i+1}
+ S_i^yS^y_{i+1}\Sigma_i^z\Sigma_{i+1}^x + S_i^zS^z_{i+1}\right)$.}
Note that $\exp(i\pi S^\alpha_j)S_j^\alpha=-S_j^\alpha$. The KT
transformation can transform a Hamiltonian into an equivalent one with a
minus sign, which indicates the FM order along either $x$- or $z$-axes,
resulting in $\mathbb{Z}_2\times\mathbb{Z}_2$ symmetry breaking. In this
case, the nonlocal string observable for $\tilde{H}_{J}$,
$O^\alpha(l,m)$ in Eq. (\ref{Osigma}), becomes the two-point
correlations $C^\alpha(l,m)$ $(\alpha=x,z)$ in Eq. (\ref{two-point})
of the transformed Hamiltonian.

Note that in terms of the KT transformation, Eq. (\ref{Ham1}) is
\mbox{$\tilde{H}_{\rm K}=-K\sum^{N/2}_{i=1}\left(S^x_{2i-1} S^x_{2i}
+ S^y_{2i} S^y_{2i+1}\Sigma_{2i}^z\Sigma_{2i+1}^x\right)$.}
This suggests that the phase diagram of the KH model with FM Kitaev
$K<0$ is similar to the one of the KH model with AFM Kitaev $K>0$ and
transformed spin correlations, i.e., $H(-K,-J)\sim H(K,J)$.
Furthermore, the spatial inversion symmetry, the time-reversal symmetry,
and the dihedral $D_2$ symmetry are preserved. In this sense, the
Haldane phase is still robust as a topological phase and protected by
these symmetries~\cite{Pol10,Pol12}. Finite correlation $C^z(1,50)$ or
finite string order parameter $O^x(1,50)$ between sites 1 and 50 on a
$N\!=\!100$ lattice with PBC indicate the FM$_z$ (LLRR) phase for
$J\!<\!J_{c1}$ ($J\!>\!J_{c1}$), see Fig. \ref{fig4}(a).
A positive (negative) sign of $O^x(1,N/2)$ for $N=60$, 100 ($N=40$, 80,
not shown) agree with the periodicity of a multiple of 4 in the LLRR
phase. When $J\!>\!0$ increases, a QPT occurs from the Kitaev QSL phase 
to the Haldane phase~\cite{Ren20}, see Fig. \ref{fig4}(b). Here finite 
$x$-correlations $\langle S^x_1S^x_i\rangle$ occur above
$J_{c3}\!\approx\!0.08$, see Fig. \ref{fig3}(d)]. The QPT at $J_{c3}$ is 
continuous and moves rightwards for increasing $N$, see the 
inset of Fig. \ref{fig4}(b). 
A more precise determination of the stability range of the QSL and of
the critical point $J_{c3}$ requires calculations on larger systems.

In summary, we characterize the ground-state properties of the
Kitaev-Heisenberg $S=1$ chain by the local and nonlocal
correlations and identify four distinct phases for $J\in(-1,1)$. 
For large negative $J$, the FM$_z$ order is favored; increasing value 
of $J$ gives a transition to an intermediate LLRR phase. These spin
correlations vanish beyond the second phase transition when $J$
approaches the Kitaev limit. In stark contrast to the gapless QSL
of the $S=1/2$ Kitaev chain, the $S\!=\!1$
chain supports a gapped QSL near $J=0$. It is characterized by the 
short-range correlations and the dimer order parameter (\ref{dimerOP}). 
Further increase of $J$ suppresses the dimer order and gives a valence 
bond solid with the singlets oriented along the $x$-direction as 
inferred from finite string order parameter $O^x$, the Haldane phase. 
It is robust against the Kitaev interactions since it is protected by 
the combination of the spatial inversion symmetry, the time-reversal 
symmetry, and the dihedral $D_2$ symmetry. It maintains its topological 
character in a range of $J>0$---and cannot evolve adiabatically to 
other phases.

The ground-state properties in the presence of anisotropy in the Kitaev
interactions and Heisenberg exchange interactions deserve further
studies. It is also an interesting future issue to investigate possible
phase transitions caused by the effect of non-Kitaev interactions, such
as off-diagonal exchange interactions, which have been extensively
investigated in a few candidate materials of the Kitaev magnets:
A$_2$IrO$_3$~\cite{Rau14}, \mbox{$\alpha$-RuCl$_3$}
\cite{Lukas2017,wang2017,Lam18,Takikawa19,Ilkem2019,Smit2020,Ran2017},
$\beta$-LiIrO$_3$~\cite{Ducatman2018,Rou17,Majumder2018,Kim2016},
K$_2$IrO$_3$~\cite{Yadav2019}.

\textit{Acknowledgements.}---This work is supported by the National
Natural Science Foundation of China (NSFC) under Grants No. 11704186
and 11474211 and by the National Science Centre (NCN, Poland)
under Project No. 2016/23/B/ST3/00839. W.-L. Y. acknowledges
support by the startup fund (Grant No. 1008-YAH20006)
of Nanjing University of Aeronautics and Astronautics.
W. C. Y. acknowledges financial support by City University
of Hong Kong through Grant No. 9610438.
A. M. Ole\'s is grateful for the Alexander von Humboldt Foundation
Fellowship \mbox{(Humboldt-Forschungspreis).}


\begin{thebibliography}{99}

\bibitem{Nus15} Z. Nussinov and J. van den Brink,
                   Compass models: Theory and physical motivations,
                   Rev. Mod. Phys. {\bf 87}, 1 (2015).

\bibitem{Aas20} D. Aasen, R. S. K. Mong, B. M. Hunt, D. Mandrus, and J. Alicea,
                   Electrical Probes of the Non-Abelian Spin Liquid in Kitaev Materials,
                   Phys. Rev. X {\bf 10}, 031014 (2020).

\bibitem{Kitaev} A. Y. Kitaev,
                   Fault-tolerant quantum computation by anyons,
                   Ann. Phys. (Amsterdam) {\bf 303}, 2 (2003);
                   Anyons in an exactly solved model and beyond,
                   Ann. Phys. (Amsterdam) {\bf 321}, 2 (2006).

\bibitem{Jackeli09} G. Jackeli and G. Khaliullin,
                   Mott Insulators in the Strong Spin-Orbit Coupling Limit:
                   From Heisenberg to a Quantum Compass and Kitaev Models,
                   Phys. Rev. Lett. {\bf 102}, 017205 (2009).

\bibitem{Dag17} S. M. Winter, A. A. Tsirlin, M. Daghofer, J. van den Brink,
                   Y.~Singh, P. Gegenwart, and R. Valenti,
                   Models and materials for generalized Kitaev magnetism,
                   J. Phys.: Condens. Matter {\bf 29}, 493002 (2017).

\bibitem{Rau14} J. G. Rau, E. Kin-Ho Lee, and H.-Y. Kee,
                   Generic Spin Model for the Honeycomb Iridates
                   beyond the Kitaev Limit,
                   Phys. Rev. Lett. {\bf 112}, 077204 (2014).

\bibitem{Juraj} J. Rusna\v{c}ko, D. Gotfryd, and J. Chaloupka,
                   Kitaev-like honeycomb magnets:
                   Global phase behavior and emergent effective models,
                   Phys. Rev. B {\bf 99}, 064425 (2019).

\bibitem{Wan20} W. Yang, A. Nocera, T. Tummuru, H.-Y. Kee, and Ian \mbox{Affleck,}
                   Phase Diagram of the Spin-1/2 Kitaev-Gamma Chain
                   and Emergent SU(2) Symmetry,
                   Phys. Rev. Lett. {\bf 124}, 147205 (2020).

\bibitem{Ian}   W. Yang, A. Nocera, and Ian Affleck,
                   Comprehensive study of the phase diagram
                   of the spin-1/2 Kitaev-Heisenberg-Gamma chain,
                   arXiv:2004.12954;
                   Spin wave theory of one-dimensional generalized Kitaev model,
                   arXiv:2007.07509.

\bibitem{Cha13} J. Chaloupka, G. Jackeli, and G. Khaliullin,
                   Zigzag Magnetic Order in the Iridium Oxide Na$_2$IrO$_3$,
                   Phys. Rev. Lett. {\bf 110}, 097204 (2013).

\bibitem{Got17} D. Gotfryd, J. Rusna\v{c}ko, K. Wohlfeld, G. Jackeli,
                   J. Chaloupka, and A. M. Ole\'s,
                   Phase diagram and spin correlations of the Kitaev-Heisenberg
                   model: Importance of quantum effects,
                   Phys. Rev. B {\bf 95}, 024426 (2017).

\bibitem{Agr18} C. E. Agrapidis, J. van den Brink, and S. Nishimoto,
                    Ordered states in the Kitaev-Heisenberg model:
                    From 1D chains to 2D honeycomb,
                    Scientific Reports {\bf 8}, 1815 (2018).

\bibitem{Brzezicki07} W. Brzezicki, J. Dziarmaga, and A. M. Ole\'s,
                    Quantum phase transition in the one-dimensional compass model,
                    Phys. Rev. B {\bf 75}, 134415 (2007).

\bibitem{Brzezicki09} W. Brzezicki and A. M. Ole\'s,
                    Quantum Phase Transition in the One-Dimensional XZ Model,
                    Acta Phys. Polon. A {\bf 115}, 162 (2009).

\bibitem{You08} W.-L. You and G.-S. Tian,
                   Quantum phase transition in the one-dimensional
                   compass model using the pseudospin approach,
                   Phys. Rev. B {\bf 78}, 184406 (2008).

\bibitem{You10} W.-L. You, G.-S. Tian, and H.-Q. Lin,
                   The low-energy states and directional long-range
                   order in the two-dimensional quantum compass model,
                   J. Phys. A: Math. Theor. {\bf 43}, 275001 (2010).

\bibitem{You11} P.-S. He,  W.-L. You, and G.-S. Tian,
                   Two-dimensional quantum compass model in
                   a staggered field: some rigorous results,
                   Chin. Phys. B  {\bf 20}, 017503 (2011).

\bibitem{Eriksson} E. Eriksson and H. Johannesson,
                   Multicriticality and entanglement
                   in the one-dimensional quantum compass model,
                   Phys. Rev. B {\bf 79}, 224424 (2009).

\bibitem{Subrahmanyam} V. Subrahmanyam,
                   Block entropy for Kitaev-type spin chains in a transverse field,
                   Phys. Rev. A  {\bf 88}, 032315 (2013).

\bibitem{Liu}   G.-H. Liu, W. Li, W.-L. You, G.-S. Tian, and G. Su,
                   \mbox{Matrix} product state and quantum phase transitions
                   in the one-dimensional extended quantum compass model,
                   Phys. Rev. B {\bf 85}, 184422 (2012).

\bibitem{Baskaran07} G. Baskaran, S. Mandal, and R. Shankar,
                   Exact Results for Spin Dynamics
                   and Fractionalization in the Kitaev Model,
                   Phys. Rev. Lett. {\bf 98}, 247201 (2007).

\bibitem{You12} W.-L. You,
                   Quantum correlation in one-dimensional
                   extended quantum compass model,
                   Eur. Phys. J. B  {\bf 85}, 83 (2012).

\bibitem{Stavropoulos19} P. P. Stavropoulos, D. Pereira, and H.-Y. Kee,
                   Microscopic Mechanism for a Higher-Spin Kitaev Model,
                   Phys. Rev. Lett. {\bf 123}, 037203 (2019).

\bibitem{Haldane1} F. D. M. Haldane,
                   Continuum dynamics of the 1-D Heisenberg antiferromagnet:
                   Identification with the O(3) nonlinear sigma model,
                   Phys. Lett. A {\bf 93}, 464 (1983).

\bibitem{Haldane2} F. D. M. Haldane,
                   Nonlinear Field Theory of Large-Spin Heisenberg Antiferromagnets:
                   Semiclassically Quantized \mbox{Solitons} of the One-Dimensional
                   Easy-Axis N\'{e}el State,
                   Phys. Rev. Lett. {\bf 50}, 1153 (1983).

\bibitem{White93} S. R. White and D. A. Huse,
                   Numerical renormalization-group study of low-lying
                   eigenstates of the antiferromagnetic $S=1$ Heisenberg chain,
                   Phys. Rev. B {\bf 48}, 3844 (1993).

\bibitem{Cohen14} I. Cohen and A. Retzker,
                   Proposal for Verification of the Haldane Phase Using Trapped Ions,
                   Phys. Rev. Lett. {\bf 112}, 040503 (2014).

\bibitem{Kennedy} T. Kennedy and H. Tasaki,
                   Hidden ${\mathbb{Z}}_{2}\times{\mathbb{Z}}_{2}$
                   symmetry breaking in Haldane-gap antiferromagnets,
                   Phys. Rev. B {\bf 45}, 304 (1992).

\bibitem{Takada} S. Takada and K. Kubo,
                   Nonlocal Unitary Transformations on $S=1$
                   Antiferromagnetic Spin Chains,
                   J. Phys. Soc. Jpn.  {\bf 60}, 4026 (1991).

\bibitem{Nijs}  M. den Nijs and K. Rommelse,
                   Preroughening transitions in crystal surfaces
                   and valence-bond phases in quantum spin chains,
                   Phys. Rev. B {\bf 40}, 4709 (1989).

\bibitem{Tasaki} H. Tasaki,
                   Quantum liquid in antiferromagnetic chains:
                   A~stochastic geometric approach to the Haldane gap,
                   Phys. Rev. Lett. {\bf 66}, 798 (1991).

\bibitem{Berg}  E. Berg, E. G. Dalla Torre, T. Giamarchi, and E. Altman,
                   Rise and fall of hidden string order of lattice bosons,
                   Phys. Rev. B {\bf 77}, 245119 (2008).

\bibitem{Hatsugai} Y. Hatsugai and M. Kohmoto,
                   Numerical study of the hidden antiferromagnetic
                   order in the Haldane phase,
                   Phys. Rev. B {\bf 44}, 11789 (1991).

\bibitem{Liu15} G.-H. Liu, L.-J. Kong, and W.-L. You,
                   Quantum phase transitions in spin-1 compass chains,
                   Eur. Phys. J. B {\bf 88}, 284 (2015).

\bibitem{Bas08} G. Baskaran, D. Sen, and R. Shankar,
                   Spin-$S$ Kitaev model: Classical ground states,
                   order from disorder, and exact correlation functions,
                   Phys. Rev. B {\bf 78}, 115116 (2008).

\bibitem{Sen10} D. Sen, R. Shankar, D. Dhar, and K. Ramola,
                   Spin-1 Kitaev model in one dimension,
                   Phys. Rev. B {\bf 82}, 195435 (2010).

\bibitem{You07} W.-L. You, G.-S. Tian, and H.-Q. Lin,
                   Existence of long-range orbital order
                   in a two-dimensional orbital-only model,
                   Phys. Rev. B {\bf 75}, 195118 (2007).

\bibitem{Hickey20} C. Hickey, C. Berke, P. P. Stavropoulos, H.-Y. Kee,
                   and S.~Trebst,
                   Field-driven gapless spin liquid
                   in the spin-1 Kitaev honeycomb model,
                   Phys. Rev. Research {\bf 2}, 023361 (2020).

\bibitem{Tro10} F. Trousselet, A. M. Ole\'{s}, and P. Horsch,
                   Compass-Heisenberg model on the square lattice
                   --Spin order and elementary \mbox{excitations,}
                   Europhys. Lett (EPL) {\bf 91},  40005 (2010);
                   Magnetic properties of nanoscale compass-Heisenberg planar clusters,
                   Phys. Rev. B \textbf{86}, 134412 (2012).

\bibitem{Chen07} S. Chen, Li Wang, S.-J. Gu, and Y. Wang,
                   Fidelity and quantum phase transition for the
                   Heisenberg chain with next-nearest-neighbor interaction,
                   Phys. Rev. E \textbf{76}, 061108 (2007).

\bibitem{Yang20} Y. Yang, S.-J. Ran, Xi Chen, Z.-Z. Sun, S.-S. Gong,
                   Z. Wang, and G. Su,
                   Reentrance of the topological phase
                   in a spin-1 frustrated Heisenberg chain,
                   Phys. Rev. B {\bf 101}, 045133 (2020).

\bibitem{Kennedy90} T. Kennedy,
                   Exact diagonalisations of open spin-1 chains,
                   J.~Phys.: Condens. Matter {\bf 2}, 5737 (1990).

\bibitem{Affleck87} I. Affleck, T. Kennedy, E. H. Lieb, and H. Tasaki,
                   Rigorous \mbox{results} on valence-bond ground states in antiferromagnets,
                   Phys. Rev. Lett. {\bf 59}, 799 (1987).

\bibitem{Oshikawa92} M. Oshikawa,
                   Hidden $\mathbb{Z}_2\times\mathbb{Z}_2$ symmetry in quantum
                   spin chains with arbitrary integer spin,
                   J. Phys.: Condens. Matter {\bf 4}, 7469 (1992).

\bibitem{Pol10} F. Pollmann, A. M. Turner, E Berg, and M. Oshikawa,
                   Entanglement spectrum of a topological phase in one dimension,
                   Phys. Rev. B {\bf 81}, 064439 (2010).

\bibitem{Pol12} F. Pollmann, E. Berg, A. M. Turner, and M. Oshikawa,
                   Symmetry protection of topological phases
                   in one-dimensional \mbox{quantum} spin systems,
                   Phys. Rev. B {\bf 85}, 075125 (2012).

\bibitem{Ren20} J. Ren, W.-L. You, and A. M. Ole\'{s},
                   Quantum phase transitions in a spin-1
                   antiferromagnetic chain with long-range interactions
                   and modulated single-ion anisotropy,
                   Phys. Rev. B {\bf 102}, 024425 (2020).

\bibitem{Takikawa19} D. Takikawa and S. Fujimoto,
                   Impact of off-diagonal exchange interactions
                   on the Kitaev spin-liquid state of $\alpha$-RuCl$_3$,
                   Phys. Rev. B {\bf 99}, 224409 (2019).

\bibitem{Lukas2017} L. Janssen, E. C. Andrade, and M. Vojta,
                   Magnetization processes of zigzag states on the
                   honeycomb lattice: Identifying spin models for
                   $\ensuremath{\alpha}\text{\ensuremath{-}}{\mathrm{RuCl}}_{3}$
                   and ${\mathrm{Na}}_{2}{\mathrm{IrO}}_{3}$,
                   Phys. Rev. B {\bf 96}, 064430 (2017).

\bibitem{wang2017} W. Wang, Z.-Y. Dong, S.-Li Yu, and J.-X. Li,
                   Theoretical \mbox{investigation} of magnetic dynamics in
                   $\ensuremath{\alpha}\text{\ensuremath{-}}{\mathrm{RuCl}}_{3}$,
                   Phys. Rev. B {\bf 96}, 115103 (2017).

\bibitem{Lam18} P. Lampen-Kelley, S. Rachel, J. Reuther,
                   J.-Q. Yan, A. Banerjee, C. A. Bridges, H. B. Cao,
                   S. E. Nagler, and D. Mandrus,
                   Anisotropic susceptibilities in the honeycomb Kitaev system $\ensuremath{\alpha}\text{\ensuremath{-}}{\mathrm{RuCl}}_{3}$,
                   Phys. Rev. B {\bf 98}, 100403(R) (2018).

\bibitem{Ilkem2019} I. O. Ozel, C. A. Belvin, E. Baldini, I. Kimchi,
                   S. Do, \mbox{K.-Y. Choi,} and N. Gedik,
                   Magnetic field-dependent low-energy magnon dynamics in
                   $\ensuremath{\alpha}\text{\ensuremath{-}}{\mathrm{RuCl}}_{3}$,
                   Phys. Rev. B {\bf 100}, 085108 (2019).

\bibitem{Smit2020} R. L. Smit, S. Keupert, O. Tsyplyatyev,
                   P. A. Maksimov, A.~L.~Chernyshev, and P. Kopietz,
                   Magnon damping in the zigzag phase of the
                   Kitaev-Heisenberg-$\mathrm{\ensuremath{\Gamma}}$ model on a honeycomb lattice,
                   Phys. Rev. B {\bf 101}, 054424 (2020).

\bibitem{Ran2017} K. Ran, J. Wang, W. Wang, Z.-Y. Dong, X. Ren, S. Bao \textit{et al.},
                   Spin-Wave Excitations Evidencing the Kitaev \mbox{Interaction} in Single
                   Crystalline $\ensuremath{\alpha}\text{\ensuremath{-}}{\mathrm{RuCl}}_{3}$,
                   Phys. Rev. Lett. {\bf 118}, 107203 (2017).

\bibitem{Ducatman2018} S. Ducatman, I. Rousochatzakis, and N. B. Perkins,
                   Magnetic structure and excitation spectrum of the hyperhoneycomb \mbox{Kitaev} magnet
                   $\ensuremath{\beta}\text{\ensuremath{-}}{\mathrm{Li}}_2{\mathrm{IrO}}_3$,
                   Phys. Rev. B {\bf 97}, 125125 (2018).


\bibitem{Rou17} I. Rousochatzakis and N. B. Perkins,
                   Classical Spin Liquid Instability Driven By
                   Off-Diagonal Exchange in Strong Spin-Orbit Magnets,
                   Phys. Rev. Lett. {\bf 118}, 147204 (2017);
                   Magnetic field induced evolution of intertwined orders in the Kitaev \mbox{magnet}
                   $\ensuremath{\beta}\text{\ensuremath{-}}{\mathrm{Li}}_2{\mathrm{IrO}}_3$,
                   Phys. Rev. B {\bf 97}, 174423 (2018).

\bibitem{Majumder2018} M. Majumder, R. S. Manna, G. Simutis, J. C. Orain,
                   T. Dey, F. Freund \textit{et al.}, 
                   Breakdown of Magnetic Order in the Pressurized Kitaev Iridate $\ensuremath{\beta}\text{\ensuremath{-}}{\mathrm{Li}}_2{\mathrm{IrO}}_3$,
                   Phys. Rev. Lett. {\bf 120}, 237202 (2018).

\bibitem{Kim2016} H.-S. Kim, Y. B. Kim, and H.-Y. Kee,
                   Revealing frustrated local moment model
                   for pressurized hyperhoneycomb iridate:
                   Paving the way toward a quantum spin liquid,
                   Phys. Rev. B {\bf 94}, 245127 (2016).

\bibitem{Yadav2019} R. Yadav, S. Nishimoto, M. Richter,
                   J. van den Brink, and~R.~Ray,
                   Large off-diagonal exchange couplings and
                   spin liquid states in ${C}_{3}$-symmetric iridates,
                   Phys. Rev. B {\bf 100}, 144422 (2019).


\end{thebibliography}
\end{document}